# Physics-Based TCAD Modeling of Nanowire Cryotrons with Coupled Superconducting and Thermal Dynamics

Abdullah All Hasib[1], Md Shafayat Hossain[2], Ahmedullah Aziz[3], and Shamiul Alam[1*]
[1]Department of Electrical and Computer Engineering, Clemson University, Clemson, SC 29634, USA
[2]Department of Materials Science and Engineering, University of California, Los Angeles, CA 90095, USA
[3]Department of Electrical Engineering and Computer Science, University of Tennessee, Knoxville, TN 37996, USA
*Corresponding Author's Email: shamiua@clemson.edu

***Abstract*— Nanowire cryotrons (nTrons) provide gate-controlled superconducting switching, current gain, and high-impedance outputs, ideal for cryogenic computing, detector readout, and superconducting–semiconductor interfaces. However, their performance stems from strongly coupled superconducting, electrical, and thermal dynamics. While existing nTron models capture parts of this physics, a calibrated device-and-circuit simulation environment that directly connects spatial superconducting dynamics with terminal response is still missing. Here, we establish a physics-based Sentaurus technology computer-aided design (TCAD) framework for nTrons that integrates time-dependent Ginzburg–Landau dynamics, distinct electron and phonon energy equations, and mixed-mode circuit simulation. Superconducting terminal-current corrections link the custom physical solver directly to the native Sentaurus electrical solution, enabling local suppression of superconductivity, heat generation and removal, current redistribution, and external-circuit response to evolve within a cohesive framework. We calibrate the framework against experimental switching characteristics at six gate currents using a single shared parameter set, achieving sub-1 µA root-mean-square (RMS) load-current error outside the switching region and switching-current differences of approximately 1 µA or less. We then utilize the calibrated framework to explore device-level trade-offs. We find that center-gate placement yields the highest switching sensitivity, while drain-side gating results in the largest post-switch resistance. Narrower channels decrease switching current yet increase resistance. Transient and bidirectional simulations uncover electrothermal hysteresis, gate-controlled switching latency, and bias-controlled recovery. Multi-gate simulations further demonstrate cooperative suppression of superconductivity, lowering the switching threshold from about 104 µA with all gates inactive to approximately 12 µA when three gates are simultaneously activated. These findings establish a unified TCAD capability for designing nTron geometry, operating conditions, and multi-input architecture while preserving the superconducting and electrothermal physics that govern device operation.**

***Index Terms*— Cryogenic electronics, electrothermal modeling, nanowire cryotron (nTron), superconducting devices, TCAD modeling, time-dependent Ginzburg–Landau (TDGL).**

## Introduction

The rapid growth of artificial intelligence, high-performance computing, quantum technologies, and space exploration is creating new demands on electronic hardware. Data centers already consume hundreds of terawatt-hours of electricity each year, and their power demand is rising rapidly as artificial-intelligence workloads increase [1]. At the same time, spacecraft, lunar systems, and planetary probes require electronics that can operate reliably under extreme temperatures and limited power budgets, often where maintaining conventional electronics inside heated enclosures carries substantial penalties in power, mass, and complexity [2]. Quantum technologies add a third challenge: superconducting processors and many of their sensors already operate at cryogenic temperatures, yet their control, readout, memory, and signal-processing electronics must still communicate efficiently with the surrounding system [3–5]. Across these very different applications, a common engineering problem lurks: how can we compute, sense, store information, and process signals with far less energy under cryogenic conditions?

Cryogenic electronics offers a compelling opportunity because cooling fundamentally alters how materials transport charge and energy. Lower thermal agitation reduces electrical noise, reduced phonon scattering can enhance carrier mobility, and suppressed thermal activation can reduce leakage in semiconductor devices [3,6]. Superconductors provide an even more dramatic regime by carrying electrical current without DC resistance within their superconducting operating limits [7,8]. These properties have already enabled superconducting quantum processors [4], ultrasensitive photon detectors [9] and specialized high-speed logic [10]. The opportunity, however, extends beyond exploiting low resistance. Useful cryogenic systems require devices that can switch, provide gain, drive subsequent stages, store information, and communicate with semiconductor electronics while adding as little heat as possible to an already constrained cooling environment. Recent interest in energy-efficient cryogenic computing reflects this broader need for new device concepts that combine low-temperature material physics with practical electronic functionality.

Josephson junctions (JJs) and superconducting quantum interference devices (SQUIDs) have demonstrated the remarkable speed and energy efficiency of superconducting electronics, with switching frequencies reaching hundreds of gigahertz [11] and switching energies below the attojoule scale [10]. At the same time, many JJ-based circuits rely on inductive elements, magnetic-flux signaling, and small output-voltage levels, which can complicate dense integration and direct communication with high-impedance semiconductor electronics [12,13]. Their limited intrinsic gain can further require additional interface circuitry between superconducting and conventional electronic stages. These challenges create a need for complementary superconducting devices that provide an independently controlled input, current gain, and sufficiently large output impedance to drive subsequent stages directly. Heater cryotrons have already shown how this functionality enables cascading between superconducting logic stages [12]. Nanowire cryotrons, or nTrons [13], extend this capability to a compact three-terminal superconducting device and offer a promising platform for logic [14,15], memory [16–19], detector readout [20,21], in-memory computing [22,23], and superconducting–semiconductor interfaces [24,25].

An nTron's performance emerges from the coupled dynamics of superconductivity, thermal, electrical, and circuit dynamics rather than from any one process alone. This coupling makes nTrons scientifically interesting and difficult to engineer. Experimental studies reveal the terminal switching characteristics, but they cannot directly show how the superconducting order, electron temperature, phonon temperature, and current distribution evolve throughout the device during switching and recovery [13]. Previous modeling has captured parts of this physics [26]. Time-dependent Ginzburg–Landau simulations have described the evolution of the superconducting state and reproduced major experimental trends, while electrothermal models have examined nonequilibrium heating and geometry-dependent operation. However, a device-design environment that directly links these spatial superconducting and thermal dynamics to terminal currents and the surrounding electrical circuit remains elusive. Such a capability would allow the same physical framework to address not only why an nTron switches, but also how geometry, input excitation, circuit loading, and multiple gates determine its engineering performance.

Here, we establish a physics-based technology computer-aided design (TCAD) framework that couples superconducting-state evolution, nonequilibrium heat flow, terminal transport, and mixed-mode circuit response. Sentaurus provides the device geometry, mesh, normal-conduction solution, and external-circuit dynamics. Concomitantly, our custom solver evolves the time-dependent Ginzburg–Landau order parameter together with separate electron and phonon energy equations. We calculate superconducting terminal-current corrections from the spatial solution and feed them directly into the native mixed-mode electrical calculation. This architecture enables local suppression and recovery of superconductivity, Joule and relaxation heating, electron–phonon energy transfer, cooling, current redistribution, and load response to evolve within one coordinated simulation framework. Our implementation also supports spatially varying superconducting properties, addressing an important limitation identified in previous nTron modeling [26]. To our knowledge, this is the first reported Sentaurus-based mixed-mode TCAD implementation for nanowire cryotrons.

We calibrate the framework against experimental switching characteristics (reported in Ref. [13]) measured at multiple gate currents using one shared parameter set. We then use the calibrated model to uncover how device geometry and operating conditions control performance. We show that gate placement governs switching sensitivity and post-switch resistance differently, while channel width establishes a tradeoff between switching threshold and resistance. Transient and bidirectional simulations reveal the electrothermal origin of hysteresis and distinguish the physics controlling gate-induced switching from that governing recovery. We further extend the framework to independently driven multi-gate nTrons and quantify how simultaneous gate excitation cooperatively suppresses superconductivity and controls the switching threshold. Collectively, these results establish a practical capability for designing nTron geometry, operating conditions, and multi-input architecture while retaining the superconducting and electrothermal physics that determine device behavior.

**nTron**

nTron is a three-terminal SC switch patterned from a thin film. A channel connects the source and drain, and a gate joins the channel through a narrow constriction known as the choke. Fig. 1 shows the device structure with three different gate positions. Niobium nitride (NbN) is widely used to make nTrons [13,27], while niobium titanium nitride (NbTiN) has also been investigated [24]. In this work, we consider an nTron device made with NbN [13]. However, the framework can also be applied to nTrons made from other materials by updating the corresponding material parameters.

Initially, both the gate and the channel remain in their SC states when a certain channel-bias current is applied. Applying a sufficiently high gate current drives the choke into its resistive state, producing Joule heating. Nonequilibrium excitations spread into the adjacent channel, generating heat. This increases the channel temperature, suppressing superconductivity and reducing the channel critical current. Once the applied channel bias current exceeds this reduced threshold, the channel also switches to its resistive state. Further Joule heating can sustain or expand that region, making the channel more resistive. The increasing channel resistance then redirects the bias current into a parallel load, producing the output signal [28] . Because the nTron output arises from a heated resistive region, completing a switching cycle requires that region to cool. Recovery occurs when heat removal becomes sufficient and the remaining channel current permits retrapping into the superconducting state. Removing the gate excitation alone may leave sufficient channel dissipation to sustain resistance [29]. Therefore, a successful reset depends on both cooling and current redistribution. Please refer to Refs. [13], [25] for more details on the design and operating principles of nTrons.

**Modeling Methodology**

We model the nTron as a superconducting conductor embedded in an external electrical circuit. Its response arises from the interplay among current injections, suppression of superconductivity, heat generation, and energy removal via cooling. To represent these processes within a Sentaurus workflow, we combine a native normal-conduction device with a custom Python solver for the superconducting order parameter and the electron and phonon temperatures. Sentaurus Structure Editor (SDE) supplies the geometry and mesh, while Sentaurus Device (SDevice) solves the normal-conduction and mixed-mode circuit equations. The custom Python solver returns superconducting terminal-current corrections through circuit model interface (CMI) elements. This division of responsibilities connects the superconducting state to circuit-level observables without replacing the native normal-conduction contribution. Fig. 2 illustrates the resulting workflow.

The calculation begins with a two-dimensional representation of a 10-nm-thick NbN film. The model assumes that the electrical and thermal quantities are uniform through the film thickness. A triangular mesh is refined around the choke, gate–channel junction, and channel constriction, where current crowding and thermal gradients are expected to be the strongest. The device dimensions and material properties are provided in Table 1 and Table 2, respectively. We select the parameters based on the documented NbN material properties [28,30,31] and prior works on nTron modeling [26]. The Ginzburg–Landau time scale follows the same critical-temperature-dependent expression as in prior works, while the effective penetration depth, local critical-temperature profile, and selected thermal coefficients were adjusted within their calibration ranges to reproduce the experimental characteristics reported in Ref. [13].

We use the time-dependent Ginzburg-Landau (TDGL) equation in our model to track the spatial superconducting order parameter and its finite relaxation time. Coupled with current redistribution and local heating, the model describes how gate-induced suppression near the choke can extend into the channel and lead to the formation of a resistive region [26,32]. Although the bath temperature is 4.2 K, the active switching region can reach a higher temperature. We initialize the simulation in an equilibrium superconducting state and use the thermal solver to calculate the evolving local electron temperature, which enters the TDGL equation along with the local critical temperature. During switching, dissipation can heat the active region toward or above its local transition temperature. As the region cools, the same coupled equations describe superconducting recovery. Conventional TDGL has its strongest microscopic justification near the transition. We therefore use it here as an effective device-level description. Its spatial electrothermal coupling and calibration against multiple experimental switching curves support studies of geometry-dependent switching and current redistribution under the examined conditions. We initialize the SC order parameter using the Ginzburg-Landau equation and the same spatial critical-temperature profile used in the transient solver [26]. This initialization avoids artificial relaxation heating that could arise from starting with a nonequilibrium order parameter. For uniform material at equilibrium, the stationary solution reduces to

$$|\psi_{\mathrm{eq}}|^2 = \max(1 - T_e/T_c, 0) \tag{1}$$

Here, $\psi$ is the superconducting order parameter, $\psi_{\mathrm{eq}}$ being its equilibrium value, $T_e$ is the electron temperature, and $T_c$ is the critical temperature. Thus, the uniform bulk reference has $|\psi|^2 \simeq 0.667$ at 4.2 K. In the nonuniform device, spatial gradients and the locally reduced critical temperature change the equilibrium amplitude near the gate.

Starting from this equilibrium, the custom Python solver evolves the dimensionless complex order parameter ($\psi = |\psi|e^{i\theta}$). Its amplitude quantifies the local strength of superconductivity, and its phase gradient generates a supercurrent. The implementation uses the zero-vector-potential TDGL equation [26]:

$$\tau_{\mathrm{GL}}\left(\frac{\partial}{\partial t} + i\frac{2\pi\phi}{\Phi_0}\right)\psi = \xi^2\nabla^2\psi + \left[1 - \frac{T_e}{T_c(\mathbf{r})} - |\psi|^2\right]\psi \equiv F(\psi, T_e) \tag{2}$$

Here, $\tau_{\mathrm{GL}}$ is the order parameter time scale, $t$ is the physical time, $\phi$ is the total scalar electrical potential, $\Phi_0 (= \frac{h}{2e}$, where $h$ is Planck's constant and $e$ is the charge of the electron) is the superconducting flux quantum, $\xi$ is the coherence length, $r$ is the spatial position, and $\theta$ is the phase of the complex state. The term $T_c(\mathbf{r})$ allows spatial variation. $\nabla$ denotes the spatial gradient, and $\nabla^2$ denotes the Laplacian. $F(\psi, T_e)$ denotes the complete dimensionless expression on the right-hand side of equation (2). The gradient term couples neighboring regions, while the temperature-dependent reaction term allows superconductivity to weaken under heating and recover during cooling. The electrical potential drives phase evolution. A natural boundary condition ($\mathbf{n} \cdot \nabla\psi = 0$) is applied at the active contacts and other terminal boundaries, where $\mathbf{n}$ is the outward unit vector perpendicular to the device boundary. Because Sentaurus TCAD treats defined contacts as resistive metals by default, this boundary condition is used to preserve superconductivity at the contacts rather than impose a normal-metal condition with $\psi = 0$.

The resulting supercurrent is combined with the normal current through the Ginzburg–Landau supercurrent density, local Ohm's law, and current-continuity equation [26]:

$$\begin{gathered} \mathrm{J}_s = \frac{\Phi_0}{2\pi\mu_0\lambda^2}\mathrm{Im}(\psi^*\nabla\psi), \\ \mathrm{J}_n = -\sigma_n\nabla\phi, \\ \nabla\cdot(\mathrm{J}_n + \mathrm{J}_s) = 0. \end{gathered} \tag{3}$$

Here, $J_{\mathrm{s}}$ and $J_{\mathrm{n}}$ are the superconducting and normal current densities, respectively. The permeability $\mu_0$ and effective penetration depth $\lambda$ set the supercurrent scale. Here, $\sigma_n$ is the normal-state electrical conductivity, and $\rho_n$ is the normal-state electrical resistivity. SDevice solves the ordinary constant-resistivity conductor and circuit at a fixed native temperature of 4.2 K. The superconducting and thermal fields are maintained by a Python solver. The transport coupling used here is the conductor current-continuity operator, rather than a semiconductor carrier-density model.

To connect the SC contribution to this native operator, the total potential is decomposed as $\phi = \phi_n + \phi_s$. The native solution ($\phi_n$) carries the terminal voltages, whereas the superconducting correction ($\phi_s$) has zero potential at every terminal. Using the normal-conductance matrix ($K_n$) exported from Sentaurus and the linear triangular shape functions ($N_i$), the correction is obtained from equation (4). Here, d is the film thickness, and $\Gamma_k$ represents terminal k. The SC correction current can be derived from the following equations-

$$K_n\phi_n = 0,$$
$$K_n\phi_s = b_s,$$
$$\text{at interior nodes,} \quad (b_s)_i = d\int_\Omega \nabla N_i \cdot \mathbf{J}_s \, dA, \quad \phi_s|_{\Gamma_k} = 0. \tag{4}$$

Here, $b_s$ is the assembled supercurrent source vector, and $(b_s)_i$ is its component at mesh node $i$. $\Omega$ is the two-dimensional device domain, and $dA$ is its differential area element. The electrical matrix is assembled from the exported native mesh coefficients, so the correction uses the normal-conduction operator. Terminal reactions then give the additional current, with positive current defined as flowing into the device:

$$\Delta I_k = \sum_{i\in\Gamma_k}(K_n\phi_s - b_s)_i\,,$$
$$I_k = I_{n,k} + \Delta I_k,$$
$$\sum_k \Delta I_k \simeq 0. \tag{5}$$

Here, $\Delta I_k$ is the superconducting terminal-current correction, $I_{\mathrm{n},k}$ is the native normal-conduction current at terminal $k$, and $I_k$ is their total. These corrections are supplied as Norton current sources between each non-source terminal and the source reference. This construction preserves the native conductance contribution, adds the superconducting response and monitors terminal current conservation.

Once the electrical fields and superconducting state are known, the thermal solver calculates where energy is deposited. The model includes normal-current Joule heating and dissipation associated with order-parameter relaxation [26]:

$$Q_J = \sigma_n|\nabla\phi|^2,$$
$$Q_{\mathrm{GL}} = \frac{1}{\mu_0\tau_{\mathrm{GL}}}\left(\frac{\Phi_0}{2\pi\xi\lambda}\right)^2 |F(\psi, T_e)|^2. \tag{6}$$

Here, $Q_{\mathrm{J}}$ is the volumetric Joule heating rate, and $Q_{\mathrm{GL}}$ is the volumetric heating rate associated with relaxation of the superconducting order parameter. Both heating terms are expressed in $W/m^3$, representing heat generated per unit volume. Joule heating is calculated using the total electrical potential, including the superconducting correction. For each connection between neighboring mesh nodes, the dissipated power is calculated as $g_{ij}(\phi_i - \phi_j)^2$, where $g_{ij}$ is the electrical conductance and $\phi_i - \phi_j$ is the potential difference between the nodes. The heat generated by relaxation of the superconducting order parameter is also transferred to the thermal mesh volumes while preserving its total power.

The deposited energy first changes the electron temperature, then exchanges with the phonon reservoir. These processes are described by the following separate equations-

$$C_e(T_e)\frac{\partial T_e}{\partial t} = \nabla\cdot[\kappa_e(T_e,T_c)\nabla T_e] + Q_J + Q_{\mathrm{GL}} - P_{eph}, \tag{7}$$

$$C_{ph}(T_{ph})\frac{\partial T_{ph}}{\partial t} = P_{eph} - \frac{u_{ph}(T_{ph}) - u_{ph}(T_b)}{\tau_{esc}}. \tag{8}$$

Here, $C_{\mathrm{e}}$ and $C_{\mathrm{ph}}$ are the electron and phonon heat capacities per unit volume, respectively, $\kappa_{\mathrm{e}}$ is the electron thermal conductivity, $P_{\mathrm{eph}}$ is the electron-to-phonon energy-transfer rate per unit volume, $T_{ph}$ is the phonon temperature, $T_b$ is the bath temperature, and $u_{ph}$ is the phonon energy per unit volume. Positive $P_{\mathrm{eph}}$ represents energy transferred from electrons to phonons. Lateral phonon diffusion is omitted. The implementation uses temperature-dependent normal-state heat capacities and a fifth-power exchange approximation:

$$C_e = \gamma T_e,$$

$$C_{ph} = \beta T_{ph}^3,$$

$$u_e = \frac{\gamma T_e^2}{2},$$

$$u_{ph} = \frac{\beta T_{ph}^4}{4},$$

$$P_{eph} = \Sigma\left(T_e^5 - T_{ph}^5\right). \tag{9}$$

Here, $\gamma$ is the electron heat-capacity coefficient, $\beta$ is the phonon heat-capacity coefficient, and $u_{\mathrm{e}}$ is the electron energy per unit volume. The electron–phonon coefficient is defined as $\Sigma = C_e(T_{ref})/[5\tau_{eph}T_{ref}^4]$, where $\tau_{\mathrm{eph}}$ is the electron–phonon relaxation-time parameter at the reference temperature ($T_{\mathrm{ref}}$). While these input times set the exchange and escape coefficients, they cannot directly predict the complete device switching or reset times. The two-temperature model permits finite-time energy exchange between electrons and phonons. Joule heating and order-parameter relaxation deposit energy in the electron reservoir, while electron–phonon exchange transfers energy between the two reservoirs with opposite signs in their energy equations [30,33]. Linear electronic and cubic phonon heat capacities approximate energy storage; their coefficients are treated as effective inputs because superconductivity, disorder, and film thickness can alter the actual values [28]. The electronic law is most directly applicable to normal regions, while the phonon law follows a Debye-type approximation [34]. The fifth-power exchange law follows established hot-electron and superconducting-nanowire models, with its coupling coefficient set by the reference electron–phonon relaxation time [35]. Lateral phonon diffusion is neglected, assuming that cooling is dominated by phonon escape into the substrate and electron heat conduction toward the contacts.

Electron heat conduction provides an additional route for removing heat from the active region. The selected thermal conductivity approximation [30,33,36–38] is

$$\kappa_e(T_e,T_c) = \frac{\mathcal{L}_0 T_e}{\rho_n}\min\left(\frac{T_e}{T_c},1\right), \qquad \mathcal{L}_0 = 2.44\times10^{-8}\ \mathrm{W\,\Omega\,K^{-2}}. \tag{10}$$

Here, $L_0$ is the Lorenz number. In the numerical update, its reduction factor is evaluated at the committed temperature and averaged symmetrically across each mesh edge, preserving pairwise heat transfer. The terminal nodes are held at $T_e = T_{ph} = T_b$, while the remaining outer boundaries have zero normal electron heat flux. Cooling therefore occurs through electron diffusion to cooled contacts and phonon escape to the substrate, with electron–phonon exchange redistributing energy internally. This energy balance explains recovery without an imposed reset rule in the spatial core. A gate-induced disturbance suppresses $|\psi|$, increases the normal-current contribution, and promotes further Joule heating. Once channel heating establishes a resistive region, removing the gate current may reduce the disturbance without extinguishing it. Recovery requires the remaining channel dissipation to fall sufficiently for diffusion and phonon escape to cool the electrons; equation (2) then permits superconductivity to re-establish.

The external circuit converts this spatial transition into an electrical output. A bias source feeds a node connected to a *10 kΩ* load resistor ($R_{Load}$) and to the device branch through an assumed channel inductance ($L_k = 1\,nH$). The circuit relations can be written as

$$I_{Bias} = I_{Load} + I_D,$$
$$V_{out} = R_{Load} \times I_{Load},$$
$$I_{Ch} = I_{Gate} + I_D,$$
$$\text{and} \quad L_k \frac{dI_{Ch}}{dt} = V_{out} - V_{core} \tag{11}$$

Here, $I_{\mathrm{Bias}}$ is the applied channel bias current, $I_{\mathrm{Load}}$ is the load current, and $V_{\mathrm{out}}$ is the voltage drop across the load. $I_D$ is the current entering the core drain, and $V_{core}$ is the drain-to-source core voltage. $I_{Ch}$ is the channel current. The gate-current term is included because the nTron gates inject current directly into the channel. As resistance develops, the circuit redirects current toward the load.

The coupled calculation follows the accepted-step sequence shown in Fig. 2. First, SDevice solves its normal-conduction and circuit equations using the previously committed Norton corrections. Its physical model interface (PMI) then transfers the accepted potential and elapsed time to a persistent Python solver. For the next update, Python combines the new normal-conduction potential with the previous superconducting potential correction as-

$$\phi_{use}^{n+1} = \phi_n^{n+1} + \phi_s^n. \tag{12}$$

Here, $\phi_{\mathrm{use}}^{n+1}$ is the potential used for the next superconducting and thermal update. Superscripts $n$ and $n+1$ identify the previously committed and newly advanced time levels, respectively. The order parameter is advanced with this potential and the committed electron temperature $T_e^n$. Equation (13) is the numerical form of the TDGL reaction–diffusion equation. It calculates the new order parameter at all mesh nodes.

$$\left[M + \Delta\tilde{t}K_\psi + \Delta\tilde{t}M\mathrm{diag}(|\psi^n|^2 - a^n)\right]\tilde{\psi} = M\psi_{rot}^n,$$
$$a_i^n = 1 - \frac{T_{e,i}^n}{T_{c,i}},$$
$$\Delta\tilde{t} = \Delta t/\tau_{\mathrm{GL}} \tag{13}$$

Here, $\Delta t$ is the physical time step. In Eq. (13), $\Delta\tilde{t}$ is the dimensionless time increment ($\Delta t/\tau_{\mathrm{GL}}$). The vector $a^n$ contains the local reaction coefficients $a_i^n$. The state $\psi_{\mathrm{rot}}^n$ is the previous order parameter after the first electrical-potential phase half-step, and $\tilde{\psi}$ is the intermediate state obtained from the reaction–diffusion solver. The matrix $M$ contains the nodal area weights, $K_\psi$ describes coupling between neighboring nodes, and the diag function applies the local reaction coefficients at each node. The nonlinear coefficient $|\psi^n|^2$ is taken from the previous step, allowing the new state to be obtained from a linear system. This approach preserves the stationary GL state and uses an iterative solver with a direct-solve fallback.

Next, the custom Python solver updates the electron and phonon temperatures by solving the numerical forms of the electron and phonon heat equations. The backward-Euler method is applied using $x = T_e^2$ and $y = T_{ph}^4$, making the stored-energy terms linear:

$$\frac{\gamma V_i}{2\Delta t}(x_i^{n+1} - x_i^n) + (D^n x^{n+1})_i + \Sigma V_i\left[(x_i^{n+1})^{5/2} - (y_i^{n+1})^{5/4}\right] = P_i \quad (14)$$

$$\frac{\beta V_i}{4\Delta t}(y_i^{n+1} - y_i^n) = \Sigma V_i\left[(x_i^{n+1})^{5/2} - (y_i^{n+1})^{5/4}\right] - \frac{\beta V_i}{4\tau_{esc}}(y_i^{n+1} - T_b^4) \quad (15)$$

Here, $V_i$ is the nodal control volume, $D^n$ describes electron heat diffusion, and $P_i$ is the combined Joule and order-parameter relaxation power. The coupled temperature equations are solved iteratively. Electron–phonon exchange transfers energy between the two reservoirs, while contact cooling and phonon escape remove energy from the core.

After updating the order parameter and temperatures, the Python solver recalculates the supercurrent, potential correction, and Norton current sources. These quantities are saved for the next electrical step. The overall coupling is sequential and first order in time. The model provides terminal currents, load voltage, and spatial distributions of $|\psi|^2$, $T_e$, and $T_{ph}$.

**Model Validation**

Our TCAD framework is calibrated and validated against experimental data at gate currents of 0, 2, 4, 6, 8, and 10 µA, reported in Ref. [13]. The experimental switching characteristics provide a reference for validating our framework's ability to reproduce gate-current-controlled channel switching, transition shape, and the resistive branch. During calibration, we adjust the effective penetration depth, electron–phonon relaxation coefficient, local choke critical temperature, and junction profile. These parameters control current capacity, heating, and the development of the resistive state. A single shared parameter set reproduces the gate-current dependence without shifting individual curves. This consistent approach supports comparison across operating conditions, while the fitted values remain effective model parameters.

Figs. 3(a)-(f) show the validation of our TCAD framework against the experimental data, which shows reasonable agreement across all the gate currents. Outside the defined switching window, the RMS current error of the load resistor is approximately 0.89–0.99 µA. Sampled switching crossings differ by less than 1 µA in bias. The experiment and our framework use the same materials, dimensions, and load resistance to have a fair comparison. We also investigate the temperature distribution at four different time instances after a gate current is applied (Fig. 3(g)). The 2D temperature distribution shows the switching of the channel.

**Device Engineering with Our TCAD Framework**

To demonstrate the capability of our TCAD framework, we investigate the effects of different gate positions, channel widths, and multi-gate configurations on device behavior. We first examine the effects of gate position on the superconducting channel. The gate position is important because the gate-induced perturbation is localized, and its position relative to the narrow channel region determines how effectively it suppresses superconductivity. Fig. 4(a) compares the channel switching characteristics for three gate configurations: source-side (Fig. 1(a)), center (Fig. 1(b)), and drain-side (Fig. 1(c)). For all three configurations, increasing the gate current reduces the channel critical current and increases the channel resistance, demonstrating the effectiveness of gate excitation in controlling the superconducting state. However, the extent of this modulation depends strongly on gate placement. At a gate current of 10 μA, the center-gate configuration exhibits the strongest suppression, with switching occurring at approximately 22 μA, compared with 38 μA and 36 μA for the source-side and drain-side configurations, respectively. This enhanced sensitivity can be attributed to the stronger influence of gate excitation on the central constricted region, where current concentration and localized heating can effectively destabilize the superconducting state. Interestingly, the resistance characteristics exhibit a different trend, with the drain-side gate producing the highest resistance, followed by the center and source-side configurations (Fig. 4(b)). The higher resistance of the drain-side gate can be attributed to the longer channel segment that carries the combined bias and gate currents. This can enhance the current concentration and Joule heating around the constriction, thereby promoting a longer resistive region and consequently increasing the post-switching resistance. This distinction indicates that the gate position influences not only the initiation of switching but also the characteristics of the resulting resistive state. Consequently, the switching threshold and the post-switching resistance should be considered separate performance metrics when optimizing the gate configuration.

Next, we examine the influence of channel width to understand how the conducting cross-section determines the switching behavior and electrical characteristics of the nTron. Fig. 4(c) shows that reducing the channel width leads to a lower channel critical current. At a gate current of 10 μA, the switching currents are approximately 26, 38, and 47 μA for channel widths of 150, 190, and 230 nm, respectively. The narrower channel therefore exhibits greater sensitivity to current-induced suppression of superconductivity. This behavior is consistent with the reduced superconducting current-carrying capacity associated with a smaller cross-sectional area, which results in a higher current density for a given applied current. The corresponding resistance characteristics further support this interpretation. The narrower channel exhibits a higher resistance than its wider counterparts (Fig. 4(d)), as expected from its smaller conducting cross-section. These results establish an important geometrical trade-off: reducing the channel width lowers the switching threshold while increasing the resistance of the resulting conducting path. Together, the gate-position and channel-width investigations provide useful design guidelines for tailoring the switching sensitivity and output characteristics of the nTron.

Having established the dependence of switching behavior on device geometry, we next investigate the reverse-bias sweep characteristics to understand the transition between resistive and superconducting states. Fig. 5(a) illustrates the hysteretic behavior of the device under a gate current of 6 μA. During the forward-bias sweep, the load current increases sharply when the bias current approaches 48–50 μA, indicating the formation of a resistive region in the channel. In contrast, during the reverse sweep, the channel remains resistive even after the bias current falls below the initial switching threshold, becoming superconducting only when the bias is reduced to approximately 12 μA. This pronounced difference between the switching and recovery thresholds demonstrates the hysteretic nature of the nTron. The underlying mechanism is associated with electrothermal feedback: initially, the superconducting channel carries current with negligible dissipation until a resistive region develops. Once this region forms, Joule heating maintains an elevated local temperature and suppresses the recovery of superconductivity. Consequently, a substantially lower bias current is required to sustain the resistive state than is required to initiate the transition. Recovery becomes possible only when the dissipated power decreases sufficiently for heat removal to dominate. This distinction between switching and recovery is an important consideration in the transient operation and reset behavior of superconducting nanowire devices.

To further understand the temporal evolution of the switching process, we perform transient simulation under an applied gate-current pulse. Fig. 5(b) presents the device response when the bias current is first increased to 70 μA, followed by the application of a 6 μA gate-current input at approximately 2.5 ns. Following gate excitation, the load current begins to increase after approximately 210 ps and approaches its high output level near 2.71 ns. This interval represents the switching delay between the application of the gate input and the development of the resistive output state under the operating conditions considered. When the gate current is removed at approximately 3.5 ns, the load current begins to decrease; however, the device does not immediately return to its superconducting state. Owing to sustained Joule heating at high bias current, the resistive region persists even after gate removal. A complete recovery occurs only when the bias current is gradually reduced, with the load current approaching zero at approximately 4.8 ns as the bias current approaches zero near 5 ns. This behavior confirms that gate excitation primarily initiates the transition, whereas recovery strongly depends on the channel's electrical and thermal conditions. Thus, removing the gate input alone is insufficient to reset the device when the applied bias maintains the resistive state.

We subsequently investigate the dependence of switching speed on gate-current amplitude to evaluate how effectively the input signal controls the temporal response of the nTron. Fig. 5(c) depicts the comparison between the switching and retrapping dynamics for different gate-current amplitudes. The results indicate that increasing the gate current progressively reduces the switching time, indicating that stronger gate excitation accelerates the formation of the resistive state. This dependence is particularly important for assessing the potential of the nTron in high-speed superconducting circuits, where the input amplitude directly influences the response latency. In contrast, the retrapping time remains unchanged because the recovery process is governed primarily by the bias-current reduction and the associated thermal relaxation. Therefore, contrary to the turn-on process, which can be accelerated through stronger gate excitation, the turn-off process requires appropriate control of the bias current to ensure complete recovery.

**Multi-Gate Simulation**

To explore the effects of multiple gates on nTron switching, we extend our TCAD framework to simulate two- and three-gate nTron devices. Unlike a conventional single-gate configuration, multiple gates allow different input signals to interact through their shared superconducting channel, providing additional control over the switching threshold [39]. This architecture is particularly interesting for input summation and reconfigurable logic [12,14], where the device response can be determined by the combination of applied gate signals rather than by individual input alone. Fig. 6(a) and Fig. 6(b) represent the two-gate and three-gate configurations.

Fig. 6(c) presents the switching characteristics of the two-gate nTron for different input combinations. When both gates are inactive (0, 0), the device switches at approximately 104 μA. Activating either gate individually with a 10 μA input significantly reduces the switching threshold to approximately 36 μA. When both gates are activated simultaneously, the threshold decreases further to approximately 27 μA, accompanied by the highest load current at a given bias. This reduction demonstrates that simultaneous gate excitation enhances the overall ability to suppress superconductivity in the shared channel.

Building on this concept, we extend the investigation to a three-gate configuration to examine the influence of an additional input and its spatial position. The three-gate configuration provides greater flexibility in controlling the channel and allows evaluation of the individual and combined contributions of the gates. As shown in Fig. 6(d), the all-off configuration exhibits a switching threshold of approximately 104 μA, akin to that of the two-gate device. However, activating the center gate individually produces an earlier transition than activating either outer gate alone, highlighting the stronger influence of gate excitation near the central constriction. Similarly, combinations involving the center gate generally produce lower switching thresholds than the combination of the two outer gates. When all three gates are activated simultaneously, switching begins at approximately 12 μA, and the load current reaches approximately 95 μA at a bias current of 120 μA. These results demonstrate that adding independently controlled gates provides greater flexibility in regulating the switching threshold, while the location of each gate determines its relative influence on the superconducting channel.

The observed behavior can be understood through the combined electrical and thermal interactions within the shared channel. Each active gate introduces a localized perturbation that can suppress superconductivity and generate additional heating in its surrounding region. When multiple gates operate simultaneously, these perturbations can interact, modifying the local current distribution and temperature profile and facilitating the formation or expansion of a resistive region. Consequently, the combined switching response is not necessarily a linear sum of the individual gate responses. Furthermore, an inactive gate may still influence the device through its electrical connection and thermal boundary conditions, thereby affecting current redistribution and heat transfer within the device.

**Discussion**

We establish a physics-based Sentaurus TCAD framework that directly connects the evolution of the superconducting state, electron–phonon heat flow, and circuit response in nTrons. Using a single calibrated parameter set, we reproduce gate-dependent switching and uncover design rules that are difficult to extract from terminal measurements alone.

Center gating yields the strongest reduction in switching current; drain-side gating produces the largest resistive-state output; and narrowing the channel lowers the switching threshold while increasing resistance. Our transient simulations further show that gate excitation controls how rapidly switching begins, whereas recovery is primarily limited by residual bias and electrothermal cooling, consistent with experimentally reported behavior.

The multi-gate results reveal an especially important opportunity. Activating three gates reduces the simulated switching threshold from approximately 104 μA to 12 μA, showing that independently driven inputs can cooperatively control the superconducting state of a shared channel. This creates a route to nTrons whose switching threshold is set by the combination and location of multiple inputs rather than by a single gate alone. Our framework therefore enables direct engineering of the quantities that determine whether an nTron can function inside a larger cryogenic system: switching threshold, output resistance, latency, recovery, and multi-input response. Extending this capability across experimentally validated geometries can enable compact multi-input logic, memory, and detector-readout circuits, as well as high-impedance interfaces between superconducting and semiconductor electronics-key building blocks for larger, more energy-efficient cryogenic computing systems.

### Author Contributions

A.A.H. and S.A. conceived the idea. A.A.H. developed custom physical models, performed the TCAD modeling and calibration, prepared the figures, and wrote the original draft. M.S.H., A.A., and S.A. analyzed the results and reviewed the manuscript. S.A. supervised the project.

### Data Availability

The data that support the plots within this paper and other findings of this study are available from the corresponding author upon reasonable request.

### Competing Interests

The authors declare no competing interests.

### *References*

[1] International Energy Agency, Energy demand from AI, in Energy and AI (2025), https://www.iea.org/reports/energy-and-ai/energy-demand-from-ai (accessed 17 November 2026).

[2] R. L. Patterson, A. Hammoud, and M. Elbuluk, Assessment of electronics for cryogenic space exploration missions, Cryogenics **46**, 231 (2006).

[3] S. J. Pauka et al., A cryogenic CMOS chip for generating control signals for multiple qubits, Nat. Electron. **4**, 64 (2021).

[4] P. Krantz, M. Kjaergaard, F. Yan, T. P. Orlando, S. Gustavsson, and W. D. Oliver, A quantum engineer's guide to superconducting qubits, Appl. Phys. Rev. **6**, 021318 (2019).

[5] J. M. Hornibrook et al., Cryogenic Control Architecture for Large-Scale Quantum Computing, Phys. Rev. Appl. **3**, 024010 (2015).

[6] H. Nyquist, Thermal Agitation of Electric Charge in Conductors, Phys. Rev. **32**, 110 (1928).

[7] J. Bardeen, L. N. Cooper, and J. R. Schrieffer, Theory of Superconductivity, Phys. Rev. **108**, 1175 (1957).

[8] A. Aziz and S. Alam, *Superconducting Memory Technologies* (Springer Nature Switzerland, 2025). doi:10.1007/978-3-031-83557-5.

[9] C. M. Natarajan, M. G. Tanner, and R. H. Hadfield, Superconducting nanowire single-photon detectors: physics and applications, Supercond. Sci. Technol. **25**, 063001 (2012).

[10] K. K. Likharev, Superconductor digital electronics, Physica C **482**, 6 (2012).

[11] W. Chen, A. V. Rylyakov, V. Patel, J. E. Lukens, and K. K. Likharev, Rapid single flux quantum t-flip flop operating up to 770 GHz, IEEE Trans. Appl. Supercond. **9**, 3212 (1999).
[12] S. Alam, M. S. Hossain, K. Ni, V. Narayanan, and A. Aziz, Voltage-controlled cryogenic Boolean logic gates based on ferroelectric SQUID and heater cryotron, J. Appl. Phys. **135**, 014903 (2024).
[13] A. N. McCaughan and K. K. Berggren, A superconducting-nanowire three-terminal electrothermal device, Nano Lett. **14**, 5748 (2014).
[14] S. Alam, D. S. Rampini, B. G. Oripov, A. N. McCaughan, and A. Aziz, Cryogenic reconfigurable logic with superconducting heater cryotron: Enhancing area efficiency and enabling camouflaged processors, Appl. Phys. Lett. **123**, 152603 (2023).
[15] S. Alam, A. N. McCaughan, and A. Aziz, Reconfigurable Superconducting Logic Using Multi-Gate Switching of a Nano-Cryotron, in *2023 Device Research Conference* (IEEE, 2023), doi:10.1109/DRC58590.2023.10186942.
[16] S. Alam, M. S. Hossain, and A. Aziz, A cryogenic memory array based on superconducting memristors, Appl. Phys. Lett. **119** (2021), doi:10.1063/5.0060716.
[17] S. Alam, M. M. Islam, M. S. Hossain, K. Ni, V. Narayanan, and A. Aziz, Cryogenic Memory Array based on Ferroelectric SQUID and Heater Cryotron, in *2022 Device Research Conference* (IEEE, 2022), doi:10.1109/DRC55272.2022.9855813.
[18] M. H. Nguyen et al., Cryogenic Memory Architecture Integrating Spin Hall Effect based Magnetic Memory and Superconductive Cryotron Devices, Sci. Rep. **10**, 248 (2020).
[19] Q. Y. Zhao, E. A. Toomey, B. A. Butters, A. N. McCaughan, A. E. Dane, S. W. Nam, and K. K. Berggren, A compact superconducting nanowire memory element operated by nanowire cryotrons, Supercond. Sci. Technol. **31**, 035009 (2018).
[20] A. N. McCaughan, Readout architectures for superconducting nanowire single photon detectors, Supercond. Sci. Technol. **31**, 040501 (2018).
[21] A. N. McCaughan, Y. Zhai, B. Korzh, J. P. Allmaras, B. G. Oripov, M. D. Shaw, and S. W. Nam, The thermally-coupled imager: A scalable readout architecture for superconducting nanowire single photon detectors, Appl. Phys. Lett. **121** (2022), doi:10.1063/5.0102154.
[22] S. Alam, J. Hutchins, M. S. Hossain, K. Ni, V. Narayanan, and A. Aziz, Cryogenic In-Memory Matrix-Vector Multiplication using Ferroelectric Superconducting Quantum Interference Device (FE-SQUID), in *2023 Design Automation Conference* (IEEE, 2023), doi:10.1109/DAC56929.2023.10247669.
[23] S. Alam, S. Thomann, S. S. Parihar, Y. S. Chauhan, K. Ni, H. Amrouch, and A. Aziz, Energy-efficient cryogenic ternary content addressable memory using ferroelectric SQUID, npj Unconv. Comput. **2**, 21 (2025).
[24] M. Tanaka, M. Suzuki, G. Konno, Y. Ito, A. Fujimaki, and N. Yoshikawa, Josephson-CMOS Hybrid Memory with Nanocryotrons, IEEE Trans. Appl. Supercond. **27** (2017), doi:10.1109/TASC.2016.2646929.
[25] A. N. McCaughan, V. B. Verma, S. M. Buckley, J. P. Allmaras, A. G. Kozorezov, A. N. Tait, S. W. Nam, and J. M. Shainline, A superconducting thermal switch with ultrahigh impedance for interfacing superconductors to semiconductors, Nat. Electron. **2**, 451 (2019).
[26] N. Yasukawa, T. Nishio, and Y. Mawatari, Ginzburg–Landau simulations of three-terminal operation of a superconducting nanowire cryotron, Supercond. Sci. Technol. **37**, 065013 (2024).
[27] A. E. Dane, A. N. McCaughan, D. Zhu, Q. Zhao, C. S. Kim, N. Calandri, A. Agarwal, F. Bellei, and K. K. Berggren, Bias sputtered NbN and superconducting nanowire devices, Appl. Phys. Lett. **111** (2017), doi:10.1063/1.4990066.
[28] J. K. W. Yang, A. J. Kerman, E. A. Dauler, V. Anant, K. M. Rosfjord, and K. K. Berggren, Modeling the Electrical and Thermal Response of Superconducting Nanowire Single-Photon Detectors, IEEE Trans. Appl. Supercond. **17**, 581 (2007).
[29] A. J. Kerman, J. K. W. Yang, R. J. Molnar, E. A. Dauler, and K. K. Berggren, Electrothermal feedback in superconducting nanowire single-photon detectors, Phys. Rev. B **79**, 100509 (2009).
[30] M. Sidorova et al., Electron energy relaxation in disordered superconducting NbN films, Phys. Rev. B **102**, 054501 (2020).
[31] A. Kamlapure, M. Mondal, M. Chand, A. Mishra, J. Jesudasan, V. Bagwe, L. Benfatto, V. Tripathi, and P. Raychaudhuri, Measurement of magnetic penetration depth and superconducting energy gap in very thin epitaxial NbN films, Appl. Phys. Lett. **96** (2010), doi:10.1063/1.3314308.
[32] L. Bishop-Van Horn, pyTDGL: Time-dependent Ginzburg-Landau in Python, Comput. Phys. Commun. **291**, 108799 (2023).
[33] A. D. Semenov, R. S. Nebosis, Y. P. Gousev, M. A. Heusinger, and K. F. Renk, Analysis of the nonequilibrium photoresponse of superconducting films to pulsed radiation by use of a two-temperature model, Phys. Rev. B **52**, 581 (1995).
[34] C. Kittel and P. McEuen, *Kittel's Introduction to Solid State Physics (8th Edition, Global Edition)* (John Wiley & Sons, Inc, 2018).
[35] A. Dane et al., Self-heating hotspots in superconducting nanowires cooled by phonon black-body radiation, Nat. Commun. **13**, 5429 (2022).
[36] F. C. Wellstood, C. Urbina, and J. Clarke, Hot-electron effects in metals, Phys. Rev. B **49**, 5942 (1994).

[37] E. M. Baeva et al., Thermal properties of NbN single-photon detectors, Phys. Rev. Appl. **10**, 064063 (2018).

[38] M. Tinkham, J. Free, C. N. Lau, and N. Markovic, Hysteretic $I - V$ curves of superconducting nanowires, Phys. Rev. B **68**, 134515 (2003).

[39] K. Zheng et al., A Superconducting Binary Encoder with Multigate Nanowire Cryotrons, Nano Lett. **20**, 3553 (2020).

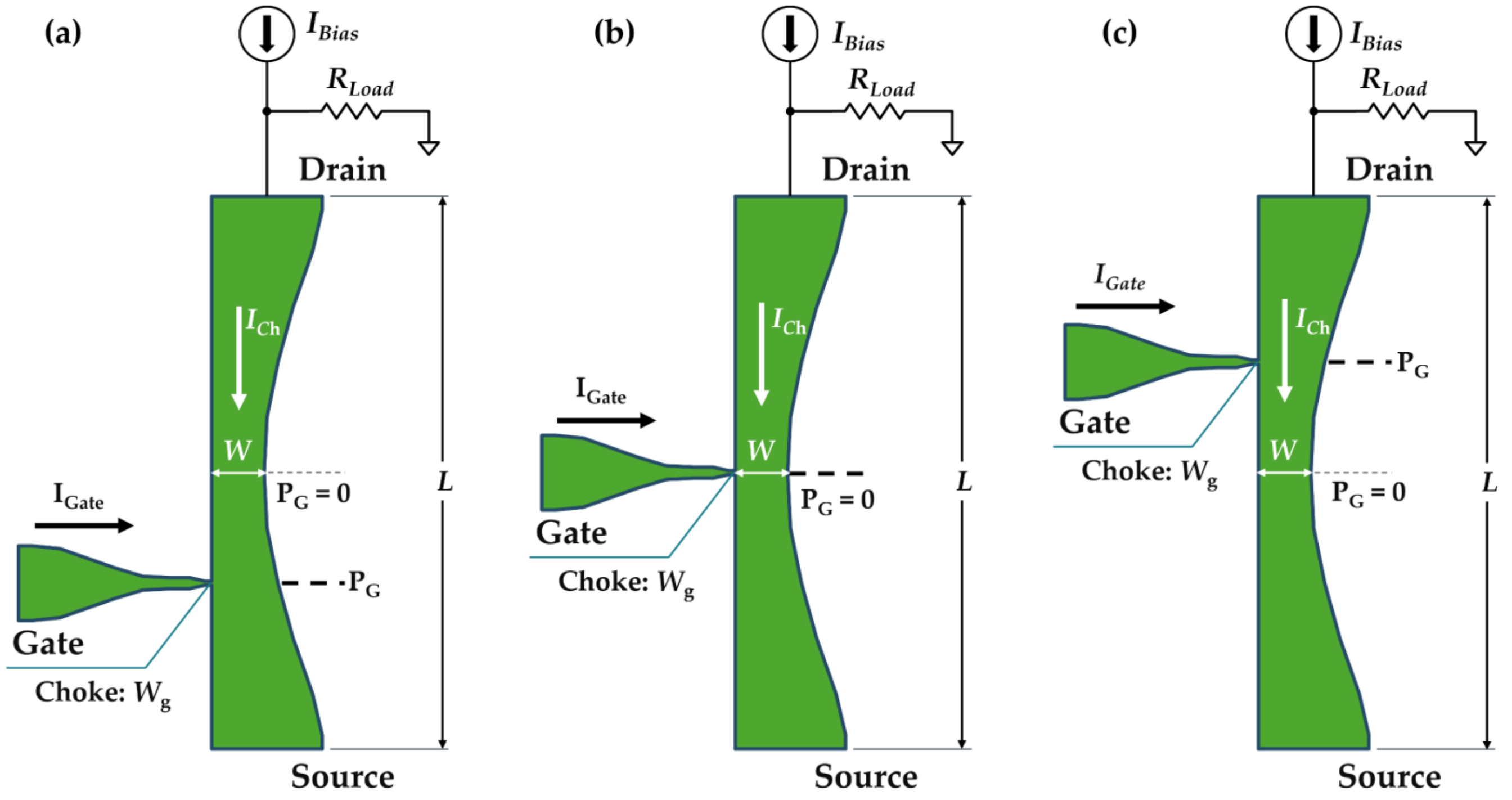

**FIG. 1 | Device structures of the nanowire cryotron (nTron).** Schematics showing nTrons with **(a)** source-side, **(b)** center, and **(c)** drain-side gate configurations. The channel length, channel width, and choke width are denoted by $L$, $W$, and $W_g$, respectively. The gate position is referenced to the channel center ($P_G = 0$). Arrows indicate the channel ($I_{Ch}$), bias ($I_{Bias}$) and gate ($I_{Gate}$) currents; $R_{Load}$ is the parallel load resistor.

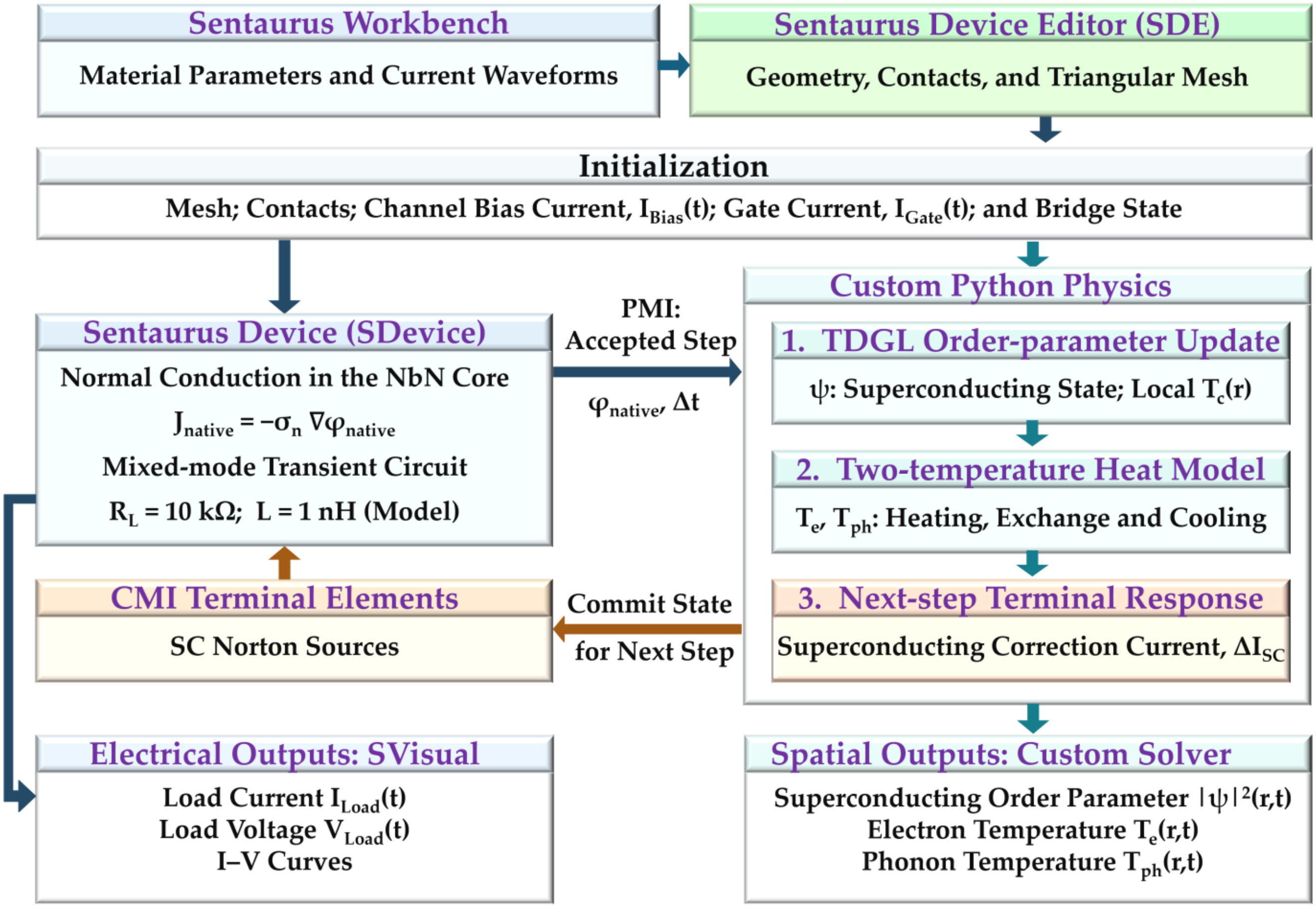

**FIG. 2 | Coupled electrothermal simulation workflow for the nTron model.** Sentaurus defines the device geometry, mesh, and mixed-mode circuit, while a custom Python solver updates the superconducting order parameter and coupled electron–phonon temperatures. Superconducting current corrections are returned to the circuit through CMI terminal elements for the next step. Outputs include terminal currents and voltages, switching characteristics, and spatial distributions of temperature and superconducting order parameter.

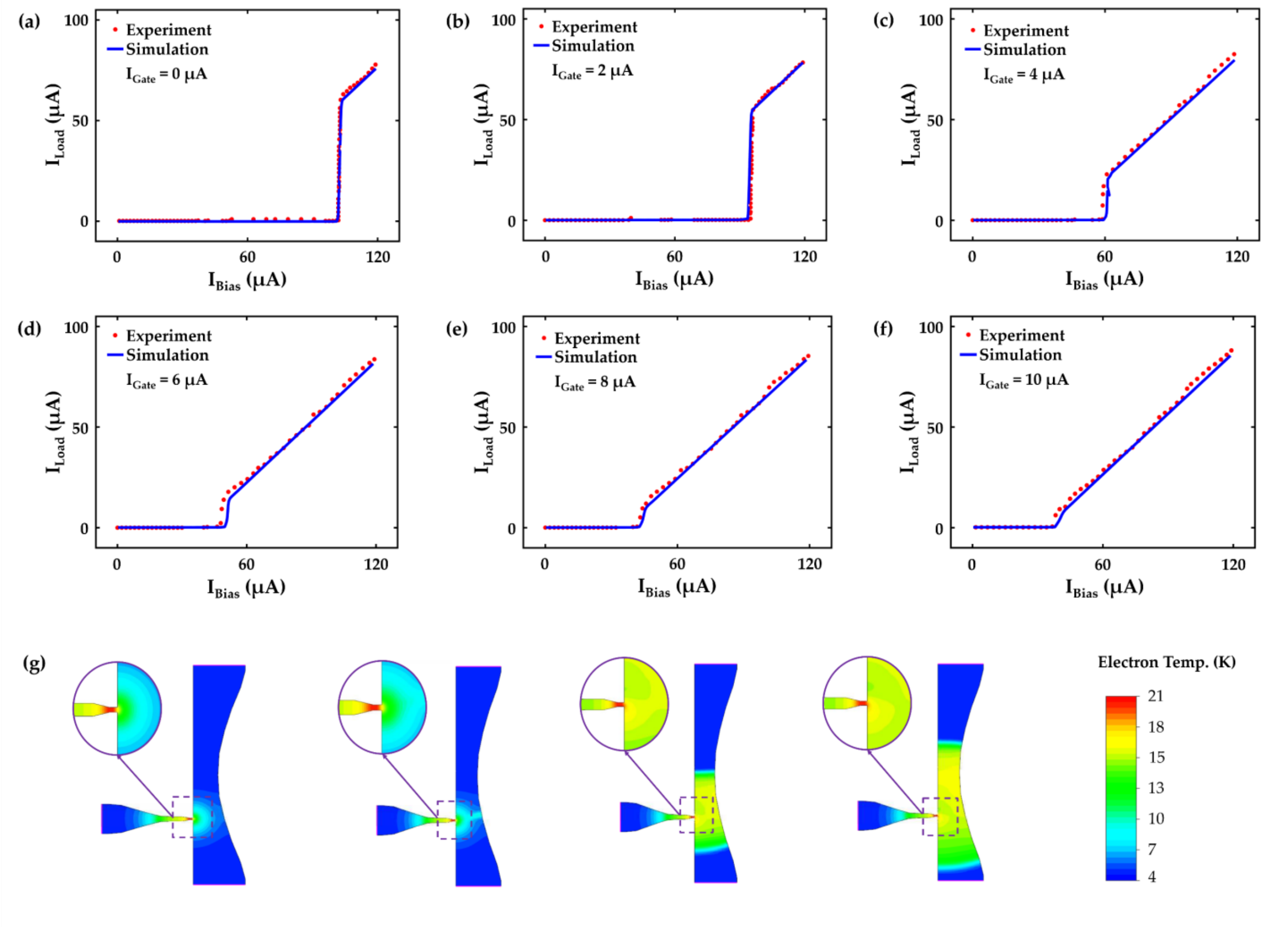

**FIG. 3 | Experimental validation of the proposed TCAD modeling framework. (a)–(f)**, Experimental (symbols) and simulated (solid lines) load current versus bias current characteristics at gate currents of 0, 2, 4, 6, 8, and 10 μA, respectively. **(g)** Simulated electron-temperature distributions illustrating localized heating near the gate choke and its expansion into the channel. The color scale indicates temperature in Kelvin. The experimental data used to validate our modeling framework is taken from Ref. [13].

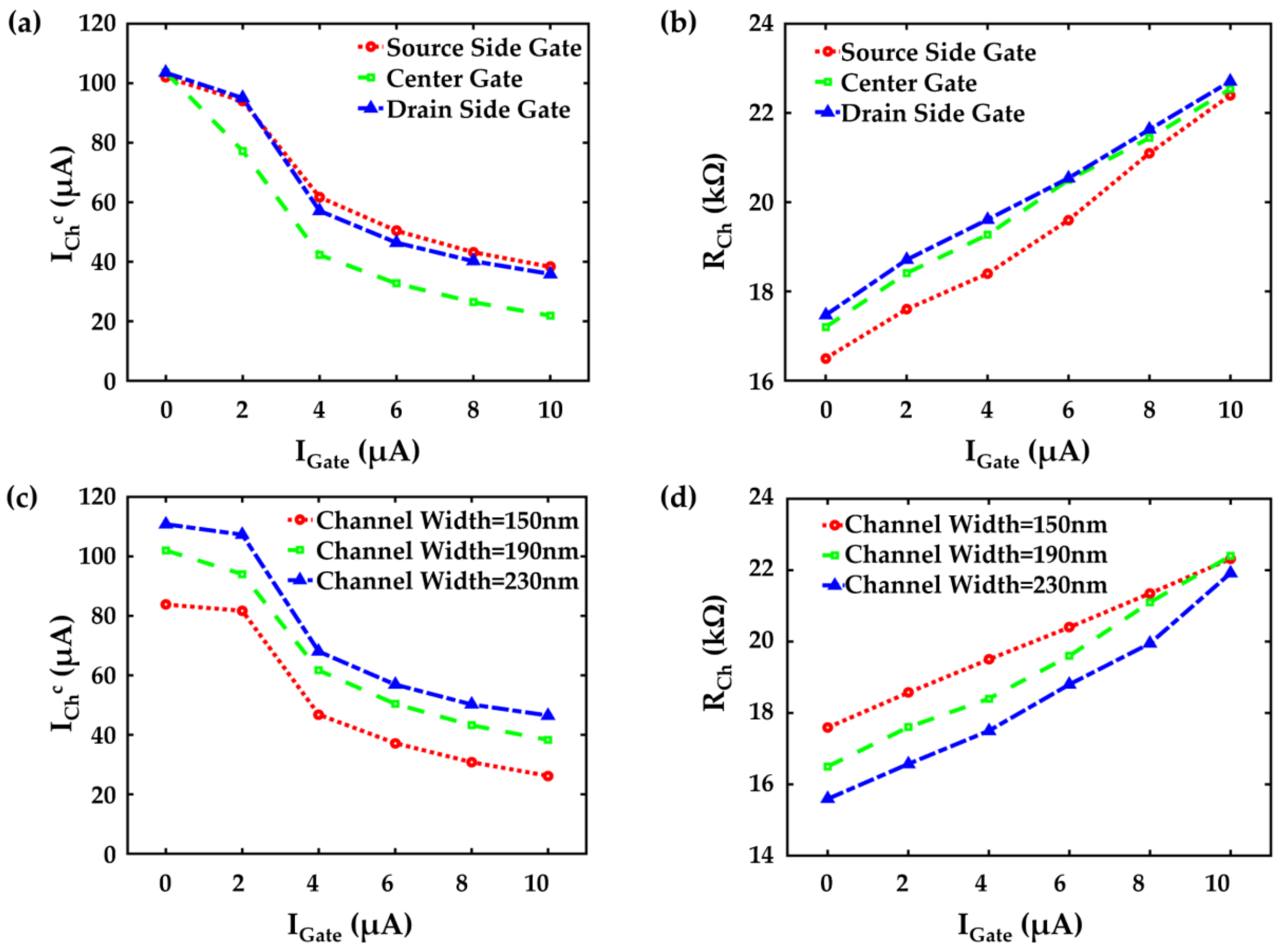

**FIG. 4 | Effects of gate position and channel width on nTron switching characteristics**. **(a)** Channel critical current ($I_{Ch}^{C}$) and **(b)** channel resistance ($R_{Ch}$) versus gate current for source-side, center, and drain-side gate positions. Effects of different channel widths on **(c)** $I_{Ch}^{C}$ and **(d)** $R_{Ch}$.

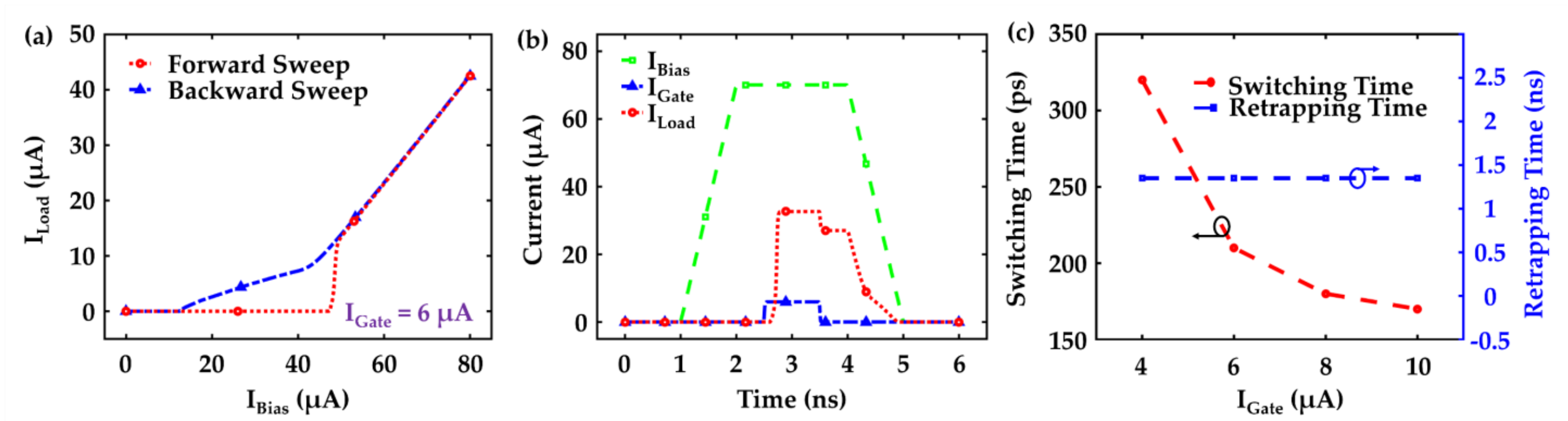

**FIG. 5 | Hysteresis and transient switching dynamics of the nTron. (a)** Load current versus bias current during forward and reverse sweeps at a gate current of 6 μA. **(b)** Time evolution of the bias, gate, and load currents during switching and recovery. **(c)** Effects of different $I_{Gate}$ on switching and retrapping times.

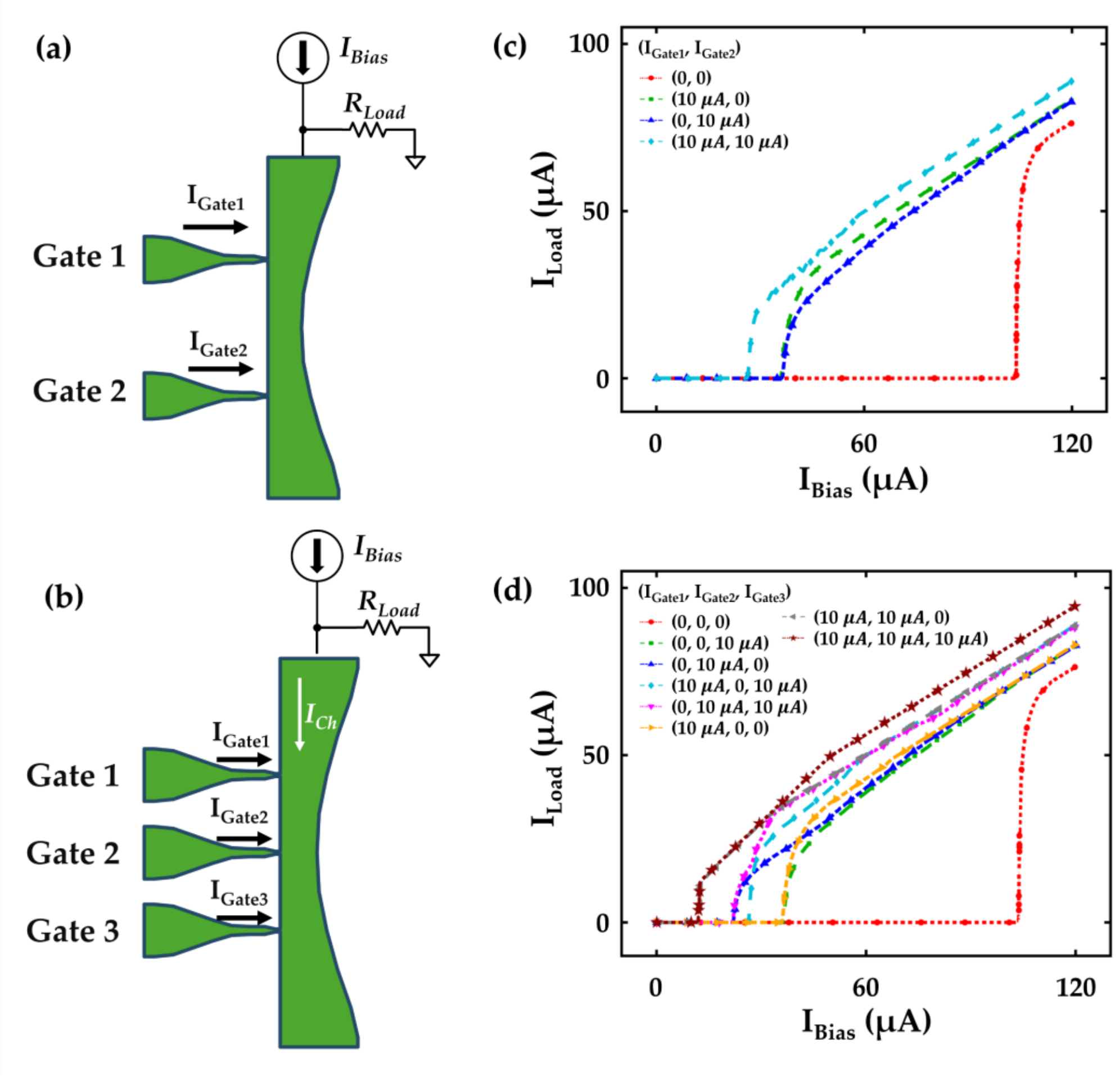

**FIG. 6 | Multi-gate nTron structures and switching characteristics.** Schematics of **(a)** two- and **(b)** three-gate devices. Load current versus bias current for combinations of independently applied gate currents of 0 and 10 μA for **(c)** two- and **(d)** three-gate devices. Combined gate excitation lowers the switching threshold, with the lowest threshold obtained when all gates are active. This behaviors match the experimental characteristics of multi-gate nTron devices reported in Ref. [39].

**TABLE 1: Device dimensions and geometry variants**

| Parameter | Value |
|---|---|
| NbN film thickness | 10 nm |
| Modeled channel length, L | 2,000 nm |
| Channel-width variants, W | 150, 190, 230 nm |
| Channel width at end contacts | 400 nm |
| Gate choke width, Wg | 15 nm |
| Reference to gate offset, PG | -400, 0, 400 nm |

**TABLE 2: Material, TDGL, thermal, and circuit parameters**

| Parameter | Value |
|---|---|
| Bath temperature, $T_b$ | 4.2 K |
| Bulk critical temperature, $T_c$ | 12.6 K |
| Coherence length, $\xi$ | 7.5 nm |
| Effective penetration depth, $\lambda$ | 352.5 nm |
| Normal-state resistivity, $\rho_n$ | $2.85 \times 10^{-6}$ Ω·m |
| Order-parameter time scale, $\tau_{\mathrm{GL}}$ | 0.238 ps |
| Electron heat-capacity coefficient, $\gamma$ | 240 J $m^{-3}$ $K^{-2}$ |
| Phonon heat-capacity coefficient, $\beta$ | 9.8 J $m^{-3}$ $K^{-4}$ |
| Reference temperature, $T_{ref}$ | 12.6 K |
| Electron–phonon relaxation time, $\tau_{eph}$ | 25 ps |
| Phonon escape time, $\tau_{esc}$ | 80 ps |
| Parallel load resistance, $R_{Load}$ | 10 kΩ |
| Channel inductance, $L_k$ | 1 nH |